\documentclass[onecolumn,usenatbib,amsmath,amssymb,amsbsy]{mn2e}
\usepackage{amsmath}
\usepackage{natbib}
\usepackage{graphicx}
\title{The Effect of Quantized Magnetic Flux Lines on the Dynamics of Superfluid Neutron Star Cores}

\begin{document}

\author[Sidery \& Alpar]{T. Sidery \& M. A. Alpar \\
FENS, Sabanc{\i} University, Orhanl{\i}, 34956 Istanbul, Turkey}

\maketitle

\begin{abstract}
We investigate dynamical coupling timescales of a neutron star's superfluid core, taking into account the interactions of
quantized neutron vortices with quantized flux lines of the proton superconductor in addition to the previously considered
scattering of the charged components against the spontaneous magnetization of the neutron vortex line.
We compare the cases where vortex motion is constrained in different ways by the array of magnetic flux tubes
associated with superconducting protons. This includes absolute pinning to and creep across a uniform array of flux lines.
The effect of a toroidal arrangement of flux lines is also considered.
The inclusion of a uniform array of flux tubes in the neutron star core significantly decreases the timescale
of coupling between the neutron and proton fluid constituents in all cases. For the toroidal component,
creep response similar to that of the inner crust superfluid is possible.
\end{abstract}

\section{Introduction}
Systems of interacting fermions are in superfluid phases at low enough temperatures. Thus the core and inner crust
regions of neutron stars consist mainly of superfluid neutrons. The neutron superfluid in the core,
carrying the bulk of the mass and moment of inertia of the star, is the most important agent in the star's rotational dynamics.
The observed timing properties, the period and spindown rate of the outer crust, are determined by the external
magnetospheric torques as well as the internal torques reflecting angular momentum exchange with the core and inner crust.
The neutron superfluid is coupled to the crust through the interactions of its quantized vortex lines with the charged components,
the protons and the electrons, which are tightly coupled to each other by electromagnetic forces and to the crust through electron viscosity.

The protons in the neutron star core are likely to be in a Type II superconducting phase with
a uniform density array of quantized flux lines carrying the magnetic flux through the core in a state of minimum free energy.
Although it makes up only a few percent of the neutron star's mass and moment of inertia, the
proton superconductor plays a crucial role in the rotational dynamics of the neutron star by defining the physical properties and
interactions of neutron vortex lines. A primary effect of the coexistence of two interacting superfluids, the neutron superfluid
and the proton superconductor is that the two superfluid currents drag each other such that the neutron superfluid
circulation around a quantized neutron vortex line drags a proton supercurrent circulating the vortex line. The induced ``spontaneous''
magnetization of the vortex lines make them scatter any relative electron and proton current strongly.
This effect leads to very short coupling times between the superfluid core and the observed outer crust of a neutron star
(Alpar, Langer \& Sauls 1984).

Observational clues and constraints on the dynamical properties of neutron star cores come from
the study of pulsar timing, pulsar glitches in particular. Glitches are observed sudden increases in the rotation frequency.
There is also a jump in the magnitude of the angular frequency derivative, the spindown rate of the pulsar.
There are no signs of change in the {\em external} magnetospheric torque, as would be reflected as changes in the electromagnetic
signature of the pulsar concurrent with the glitch. This in turn is assumed to indicate that the spin-up of the crust is due to an
exchange of angular momentum {\em within} the star. The timescales found in the relaxation of the system after a glitch extend to the
order of days and months, timescales too long for normal fluid viscosity Ekman processes. This is the argument,
based on observations, for the presence of superfluid components in neutron stars. Fractional changes in the spindown rate at glitches
and during postglitch relaxation indicate that the superfluid component involved contains about a few percent of the star's moment of inertia.
This correlates very well with the inertia of the neutron superfluid in the crustal region. The dynamical properties of the inner crust
superfluid, modeled in detail, explain the observed postglitch relaxation times (Alpar et al. 1993). The implication is that superfluids in
the neutron star's bulk, the core, must be already coupled rigidly to the observed outer crust on timescales shorter than the glitch
rise times, which have been resolved down to about 30 seconds in the Vela pulsar (Dodson, Lewis \& McCulloch 2007).
Calculations of the dynamical coupling times between
a neutron superfluid in the core and the crust and charged components of the star, based on the scattering of charges against
spontaneously magnetized neutron vortex lines, indeed yield such short relaxation
times (Alpar, Langer \& Sauls 1984, Alpar \& Sauls 1988, Andersson, Sidery \& Comer 2006, Sidery 2009).

The vortex lines in the core superfluid should also be interacting with the flux lines carried by the proton superconductor.
Since vortex lines are aligned with the rotation axis and flux lines along the magnetic axis of the neutron star, it is
unavoidable that vortex line motions, in particular motions radially outward from the rotation axis that should take place if the
core neutron superfluid is to spin down, will encounter flux lines. Vortex line flux line junctions are energetically favorable,
as the condensation energy cost of the normal matter phase in the central regions of both types of lines will be reduced by utilizing
the overlap volume at a junction. This potential pinning of vortex lines to flux lines, and modes of vortex line motion against
flux line pinning must clearly play a role in determining the rotational dynamics of the neutron star core, as was first pointed out
by Sauls (1989) as a possible agent for pulsar glitches. The vortex line flux line interaction was also proposed as an agent for
correlated spindown and magnetic field decay on evolutionary timescales for neutron stars in binaries, evolving towards the low
mass X-ray binary and subsequent millisecond pulsar configuration (Srinivasan et al 1990).

When investigating the internal dynamics of neutron stars the various interactions of the two superfluid constituents may be
of comparable importance. In this paper we investigate the dynamics of the neutron superfluid in the star's core by taking into
account, for the first time, the vortex line-flux line interaction together with the previously investigated effect of scattering
of charged particles against the spontaneously magnetized vortex lines.
We treat the neutron star core as a two component system, with superfluid neutrons as one component and
the charged particles, electrons and superconducting protons that are electromagnetically
rigidly coupled together as the second component.
If the charged fluid is treated in similar fashion as the normal component in a terrestrial superfluid,
then comparisons can be made to the dynamics of vortex lines in superfluid Helium.
In contrast to Helium the second component in the neutron star contains magnetic flux tubes.
The interactions between the neutron vortices and the magnetic flux tubes will complicate the dynamics.

There will be several possible scenarios.
The first is that the energy needed for a neutron vortex to overcome the favourable conditions at a junction and pass through a flux line
will be so large that a vortex line effectively never passes through a flux line.
When the neutron vortices' motion is thus locked to another component of the star vortices are said to be pinned.
Because we are considering pinning on flux {\em lines} - prohibition of motion through flux lines - there is still freedom to move
in the direction parallel to the magnetic flux lines.
This is different from the situation often considered in the inner crust superfluid where the neutron vortices pin to the crustal lattice.
In this case the vortices are pinned to specific fixed nuclei and so the vortex has no freedom to move in any direction unless it unpins.
At the other extreme the energy barrier of the obstruction could be so low that the neutron vortices simply pass through the
flux tubes with no hindrance. In between these two cases is that in which the vortices may gain enough energy to pass through via thermal fluctuations,
supporting some average rate of vortex motion in the direction perpendicular to the flux lines.
Thermally activated vortex motion in the presence of pinning barriers
is known as vortex creep (Alpar et al 1984a).

In Section 2 we review the dynamics in the presence of drag forces on neutron vortices alone, setting up the
description of vortex motion in response to local forces as well as the ensuing macroscopic dynamics. In Section 3,
we investigate the case of magnetic flux lines acting as absolute barriers to obstruct vortex motion.
The case of vortex creep across the flux lines is treated in Section 4. In both cases, vortex motion in the
direction parallel to the flux lines is governed by the damping force due to the scattering of charged
particles against the vortices. We consider the creep of vortex lines through a toroidal distribution of
flux lines in Section 5.

We use a simple model in which the superfluid constituents in the star, as well as the normal electron fluid rotate rigidly,
and the rotation axes of all the constituents and the external torques on the star are aligned. The magnetic flux lines
corotate with the charged component (the electrons + the superconducting protons).
The present paper addresses the dynamics of spindown (or spinup) of the neutron star under external torques, leaving the case of
precession to a subsequent paper.

\section{Dynamics with only Drag Forces on Neutron Vortex Lines}

\subsection{Local forces}
Before considering the global dynamical picture one needs to calculate the motion of a single neutron vortex under various forces.
The equation of motion for a vortex line under a force {\bf F} per unit length is the Magnus equation
\begin{equation}
\label{eq:MagnusForce}
    {\bf F} = \rho_{n} {\bf \kappa} \times \left( {\bf v}_{n} - {\bf v}_{L} \right).
\end{equation}
The kinematic right hand side of this equation is sometimes called the ``Magnus force''. Here $\rho_{n}$ is the mass
density of the superfluid neutrons, ${\bf \kappa}$ is the vorticity vector directed along the vortex line, parallel to the rotation
axis, with magnitude $\kappa = h/2m_n$ , the quantum of vorticity in terms of the Planck constant and the bare neutron mass.
${\bf v}_{n}$ and ${\bf v}_{L}$ are the velocities of the neutron superfluid and the vortex line respectively.

The physical drag force on unit length of vortex line, arising from the scattering of charged particles against the magnetization
of the neutron vortex lines has the form:
\begin{equation}
    {\bf F} = C \left( {\bf v}_{c} - {\bf v}_{L} \right) = \rho_{c} n_{v}^{-1}(\tau_v)^{-1}\left( {\bf v}_{c} - {\bf v}_{L} \right).
    \label{eq:notes011}
\end{equation}
The drag coefficient $C$ is expressed in terms of $\rho_{c}$, the mass density of the charged component,
$n_{v} = 2\Omega_n / \kappa$, the number of
vortex lines per unit area in the neutron superfluid rotating at the (uniform) rate $\Omega_n$,
and $\tau_v$ a relaxation time describing the vortex line-charged component interaction.
The velocity of the charged component is denoted by ${\bf v}_{c}$ .

Substituting this drag force in the Magnus Equation, we obtain:
\begin{equation}
\label{eq:forceBalance1}
   C \left( {\bf v}_{c} - {\bf v}_{L} \right) =
   \rho_{n} {\bf \kappa} \times \left( {\bf v}_{n} - {\bf v}_{L} \right).
\end{equation}
Eliminating ${\bf v}_L$ we obtain the force in terms of the velocity lag $({\bf v}_n - {\bf v}_c)$:
\begin{align}
  {\bf F} & = \rho_{n} \kappa \beta' \hat{{\bf \kappa}} \times ({\bf v}_{n} - {\bf v}_{c})
  + \rho_{n} \kappa \beta \hat{{\bf \kappa}} \times \hat{{\bf \kappa}} \times ({\bf v}_{n} - {\bf v}_{c}) \nonumber \\
\label{eq:MutualFriction}
            & = \rho_{n} \kappa \beta' \hat{{\bf \kappa}} \times ({\bf v}_{n} - {\bf v}_{c})
  - \rho_{n} \kappa \beta ({\bf v}_{n} - {\bf v}_{c})
\end{align}
where the last equation holds when, as in the present model, $({\bf v}_{n} - {\bf v}_{c})$ is perpendicular to $\kappa$.
The coefficients $\beta$ and $\beta'$ are introduced to make contact with the conventional notation, employed in work on
superfluid helium (ref.), to describe the strength of the dissipative (antiparallel to ${\bf v}_n$) and inertial
(along $\pm \hat{\kappa} \times \hat{v}_n $) parts of the force per unit length of a vortex line. In terms of $C$ and $\rho_{n} \kappa$
these coefficients are
\begin{equation}
  \beta = \frac{C/\rho_{n} \kappa}{1 + \left( C/ \rho_{n} \kappa \right)^{2}} \qquad \qquad \beta'
  = \frac{ \left( C/ \rho_{n} \kappa \right)^{2}}{1 + \left( C/ \rho_{n} \kappa \right)^{2}}.
\end{equation}
\\
Superfluids with $C_{strong}/\rho_{n} \kappa >> 1,\; \beta << 1,\; \beta' \cong 1$ are said to be in the ``strong coupling'' regime, while
$C_{weak}/\rho_{n} \kappa << 1,\; \beta' \cong \beta^2 << 1$ are in the weak coupling regime.

In cylindrical coordinates with ${\bf \kappa}$ and the angular velocities of the neutron superfluid
as well as the proton-electron component along the z-axis, the superfluid velocity is $v_{n,\phi}= r \Omega_n$, in the azimuthal direction.
We denote the radial and azimuthal components of the vortex line velocity ${\bf v}_{L}$ with $v_r$ and
$v_{\phi}$ respectively. All vortex lines at a given radius will have the same velocity ${\bf v}_{L}$, which
does not depend on $\phi$. Eliminating $v_{\phi}$ in the radial and axial components
of Eq.(\ref{eq:forceBalance1}), we obtain:
\begin{equation}
\label{eq:lineVelFree1}
v_r = \left[ \frac{\rho_n}{\rho_c} 2\Omega_n \tau_v + \frac{\rho_c}{\rho_n}\frac{1}{2\Omega_n \tau_v} \right]^{-1}
(\Omega_n - \Omega_c) r = \left[ \frac{\rho_n \kappa}{C} + \frac{C}{\rho_n \kappa} \right]^{-1}
(\Omega_n - \Omega_c) r = \frac{\rho_c}{\rho_n} \frac {\beta^2}{\beta'} \frac{1}{2\Omega_n \tau_v} (\Omega_n - \Omega_c) r .
\end{equation}

In this paper we shall use the results of Alpar, Langer \& Sauls (1984) who calculated the microscopic
coupling time $\tau_v$ for the scattering of electrons (and the protons which
are tightly coupled to the electrons by electromagnetic interactions) by neutron vortex lines.
The scattering is due to the very strong localized magnetic field structure carried by the neutron vortex lines.
This ``spontaneous magnetization'' arises from proton supercurrents circulating the vortex line, dragged along with
the usual neutron supercurrent circulation.
This mechanism, a two-superfluid effect, gives an extremely short coupling time
\begin{equation}
  \tau_v \cong 6.2 P (\rho_{c,14})^{-1/6} \left( \frac{m_p}{\delta m_p} \right)^2 \left( 1- \frac{\delta m_p}{m_p}\right)^{1/2}
  \cong (10-200) P
\label{eq:tauvortP}
\end{equation}
where $P$ is the neutron star's rotation period, $\rho_{c,14}$ is $\rho_{c}$ in units of $10^{14}$ gm cm$^{-3}$,
$m_p$ and $\delta m_p$ are respectively the proton mass in vacuum and
its deviation from the vacuum value in the neutron star medium,
with the effective proton mass given by ${m_p}^{*} = m_p - \delta m_p$ (Alpar, Langer \& Sauls 1984, Alpar \& Sauls 1988).
The prefactor (10-200) describes the variation of $\tau_v$ with density, proton
fraction and proton effective mass throughout the neutron star core, with shorter coupling times at higher densities.
The proton effective mass is not known very accurately, theoretical calculations giving
${m_p}^{*}/m_p \sim 0.5-0.9$. We have used the results of (Chamel \& Haensel 2006, Baldo \& Ducoin 2009)
at nuclear matter density
$\rho_{nuc} \cong 2.8 \times 10^{14}$ gm cm$^{-3}$ and at $3 \rho_{nuc}$.
The range (10-200) P also reflects the variations in the calculated parameters as reported in these two papers.
Eq. (\ref{eq:tauvortP}) gives
\begin{equation}
  \frac{C}{\rho_n \kappa} = \frac{\rho_c}{\rho_n}\frac{1}{2\Omega_n \tau_v} \sim \frac{\rho_c}{\rho_n}\frac{1}{4\pi (10-200)}
  \sim O(4 \times 10^{-4} - 2 \times 10^{-5})
\label{eq:weakcoupling}
\end{equation}
since the charged component makes up only a
small fraction of the density in the neutron star core, $\rho_c / \rho_n \approx 5 \times 10^{-2}$.
Thus the charge scattering by spontaneous magnetization of the neutron vortex lines leads to short coupling times,
which are nevertheless in the weak coupling regime in the sense of Eq. (\ref{eq:weakcoupling}).
The coefficient in Eq. (\ref{eq:lineVelFree1}) simplifies to
\begin{equation}
\label{eq:lineVelFree2}
  v_r = \frac{\rho_c}{\rho_n}\frac{1}{2\Omega_n \tau_v} (\Omega_n - \Omega_c) r
  = \frac{C}{\rho_n \kappa} (\Omega_n - \Omega_c) r.
\end{equation}

The neutron star core superfluids would be in the strong coupling regime
if much stronger coupling mechanisms existed between the charges and the neutron
vortex lines (Sedrakian \& Sedrakian 1995). Strong coupling would keep the neutron vortex lines almost in corotation
with the
charged component and flux lines. The presence of flux lines therefore should not alter the dynamical
coupling times significantly. The dynamical coupling times obtained from the particular range of
drag coefficients $C_{strong} = \eta$ of Sedrakian \& Sedrakian (1995) are much longer than the observed glitch rise times.
We will derive general expressions for dynamical coupling times
in terms of the parameters $\beta$ and $\beta'$ which can be used for both weak and strong coupling. The model of
Alpar, Langer \& Sauls (1984), which is in the weak coupling regime, will be used for the applications throughout the paper.

\subsection{Macroscopic Dynamics}

We are now ready to introduce equations for $\Omega_n$ and $\Omega_c$, the two macrosopic dynamical variables of the rotational dynamics.
The superfluid is able to spin up or spin down only by radial motion of the vortex lines,
\begin{equation}
\label{eq:spinDownSimple}
\dot{\Omega}_{n} = - \frac {n_v \kappa v_r}{r} = - \frac {2\Omega_{n} v_r}{r}.
\end{equation}
This equation, appropriate for axial symmetry about the rotation axis,
follows from the fact that a superfluid's vorticity resides in quantized vortex lines, whose number density is conserved:
\begin{equation}
\label{eq:vortexcurrent}
\dot{\Omega}_{n} = - \Omega_{n} \nabla \cdot {\bf v}_{L}
\end{equation}
When integrated over a cylindrical volume $\pi r^{2}$ times a unit length, Gauss' Theorem leads to
\begin{equation}
\label{eq:spinDownSimpleInt}
\dot{\Omega}_{n} = - \frac{\Omega_{n}}{\pi r^{2}} \oint v_{r} d l
\end{equation}
where the line integral is on a circle of radius $r$ around the equator of the star in the $x$-$y$ plane
with the $z$-axis parallel to the axis of rotation.
Eq. (\ref{eq:spinDownSimple}) follows with $d l= r d \phi$ and axial symmetry.

The last equality holds in the case of uniform rotation. Substituting the model result in Eq. (\ref{eq:lineVelFree2}) for $v_r$ gives
\begin{align}
\label{eq:nSpindownFree}
\dot{\Omega}_{n} & = - \frac{\rho_c}{\rho_n} \frac{\beta^2}{\beta'} \frac{1}{\tau_v}(\Omega_n - \Omega_c) \nonumber \\
                 & \equiv - \frac{\omega}{\tau_n}
\end{align}
where
\begin{equation}
\tau_n = \frac{\rho_n}{\rho_c}\frac{\beta'}{\beta^2}{\tau_v}
= \left[ \frac{\rho_n \kappa}{C} + \frac{C}{\rho_n \kappa} \right]\frac{P}{4\pi}
\end{equation}
and $\omega \equiv \Omega_n - \Omega_c$ denotes the lag between the rotation rates of the two components.
This is complemented by the angular momentum evolution of a two-component star under an external torque $N_{ext}$,
\begin{equation}
\label{eq:totalSpindown}
  I_n \dot{\Omega}_{n} + I_c \dot{\Omega}_{c} = N_{ext} \equiv I \dot{\Omega}_{\infty}
\end{equation}
where $I = I_n + I_c$ is the total moment of inertia and $\dot{\Omega}_{\infty} = N_{ext}/I$ denotes the steady
state spin down
(or spin up) rate reached asymptotically - hence the suffix ${\infty}$ - when both components are spinning down (or up)
at the same rate.
In the present linear drag force model, Eq.(\ref{eq:notes011}), initial conditions away from the steady state
actually relax
exponentially to the steady state. Eq.(\ref{eq:totalSpindown}) can be written, at each radius $r$ in
an axisymmetric model, as
\begin{equation}
\label{eq:totalAngMomR}
  \rho_n \dot{\Omega}_{n} + \rho_c \dot{\Omega}_{c} = \rho \dot{\Omega}_{\infty}
\end{equation}
where $\rho = \rho_n + \rho_c$ is the total density. In the neutron star core matter is mostly neutrons, with $\rho_c/\rho$
of the order of a few per cent.
From Eqns. (\ref{eq:nSpindownFree}) \& (\ref{eq:totalAngMomR}) we obtain the solution for the lag in rotation rate between
the two components
\begin{equation}
\label{eq:lagSolSimp}
  \omega(t) \equiv \Omega_n(t) - \Omega_c(t) = [\omega(0) - \omega_{\infty}] \exp (- \;t/\tau) + \omega_{\infty}
\end{equation}
where the dynamical relaxation time $\tau$ is
\begin{equation}
\label{eq:scaleTauNTau}
  \tau = \frac{\rho_c}{\rho}\tau_n = \frac{\rho_c}{\rho}\left[ \frac{\rho_n \kappa}{C} + \frac{C}{\rho_n \kappa} \right]\frac{P}{4\pi}
= \frac{\rho_n}{\rho}\frac{\beta'}{\beta^2} \tau_v \cong \tau_v,
\end{equation}
and the steady state lag is given by
\begin{equation}
\label{eq:scaleTauNLag}
  \omega_{\infty} = |\dot{\Omega}_{\infty}|\tau_n = |\dot{\Omega}_{\infty}|\tau_v \frac{\rho_n}{\rho_c}.
\end{equation}

The solutions for $\Omega_n(t)$ and $\Omega_c(t)$ for any initial conditions can be obtained from the solution
for the lag, Eq.(\ref{eq:lagSolSimp}),
and the integration of Eq.(\ref{eq:totalAngMomR}). These solutions are also characterized by
exponential relaxation with the dynamical relaxation time $\tau$.
We note that the result of Alpar \& Sauls (1988) for the dynamical relaxation time, $\tau = \rho_n\;/\;\rho_c \tau_v$, was an artefact of
keeping $\Omega_c$ constant (Andersson, Sidery \& Comer 2006).

The physics of any particular model of superfluid - normal matter coupling is described
by an equation like Eq.(\ref{eq:spinDownSimpleInt}).
When the dependence on the lag $\omega$ is linear the coupling is described by $\tau_n$. The dynamical relaxation time
$\tau$ and the steady state lag $\omega_{\infty}$ scale with $\tau_n$, as given in
Eqs. (\ref{eq:scaleTauNTau}) and (\ref{eq:scaleTauNLag}).
These generic relations hold in two-component models with different physics, as long as
the coupling is linear like in Eq.(\ref{eq:nSpindownFree}) and angular momentum evolution is given by Eq.(\ref{eq:totalSpindown}).

\section{Dynamics with Vortex Line - Flux Line Pinning and Drag Forces}

\subsection{Local forces}

Let us now introduce flux lines. The proton flux lines act as barriers offering obstruction to neutron vortex motion
in the direction perpendicular to the flux lines.
\begin{figure}
    \label{fig:vortexLine}
    \centering
      \includegraphics[width=8cm]{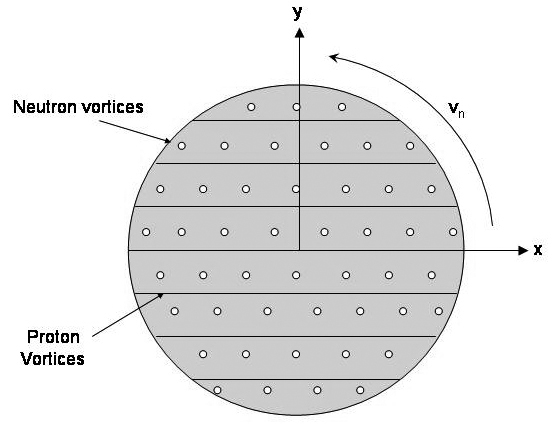}
    \caption{A sketch of the neutron star core. In the frame fixed with the charged fluid the magnetic vortices are stationary and
    parallel to the $x$-axis. The neutron vortices can move in the $x$ direction, encountering only the drag force.}
\end{figure}
To impose this we introduce a cartesian coordinate system in which the $z$-axis is parallel to the axis of rotation
(See Figure (\ref{fig:vortexLine})).
We fix the $x$-axis in the direction of the proton vortices (projected on the $x$-$y$ plane) and so the coordinates will
rotate with the proton fluid.
The neutron vortices can move in the $x$ direction, moving parallel to the flux lines, but are obstructed
in the $y$ direction,
perpendicular to the flux lines.

To include the obstruction by the flux lines we use a directionally dependent drag coefficient
and rewrite the balance of forces as
\begin{equation}
\label{eq:notes012}
    {\bf F} = C_{x}\left( v_{x,c} - v_{x,L} \right)\hat{{\bf x}} +  C_{y}\left( v_{y,c} - v_{y,L} \right)\hat{{\bf y}} =
    \rho_{n} {\bf \kappa} \times \left( {\bf v}_{n} - {\bf v}_{L} \right)
\end{equation}
Noting that this still holds in a rotating frame we can set $v_{x,c} = v_{y,c} = 0$ by using the coordinate system
fixed to the proton flux lines, proton superconductor and electrons (the charged, $c$, component).
The $x$ and $y$ components of this equation are
\begin{align}
  C_{x} v_{x,L} &= \rho_{n} \kappa \left( v_{y,n} - v_{y,L} \right)
  \label{eq:notes013}\\
  C_{y} v_{y,L} &= - \rho_{n} \kappa \left( v_{x,n} - v_{x,L} \right).
  \label{eq:notes014}
\end{align}
The case of absolute pinning in the $y$ direction corresponds to the limit $C_{y} \to \infty$ .
Equation (\ref{eq:notes014}) shows that for a reasonable solution we must have $v_{y,L} = 0$.
To put it more carefully, the physical force in the $y$ direction is not necessarily a linear drag force,
it is some probably highly nonlinear force that imposes $v_{y,L} = 0$, so that Eq. (\ref{eq:notes014}) becomes
\begin{equation}
  F_{y} = \rho_{n} \kappa \left( v_{x,n} - v_{x,L} \right).
\label{eq:carenotes014}
\end{equation}
As the vortices are not constrained by the flux lines in the $x$ direction $C_{x}$ is simply the
electron scattering drag coefficient $C$.
Substituting this into Eq.(\ref{eq:notes013}) gives us the force on unit length of vortex line as
\begin{equation}
F_{x} =  - C v_{x,L} = - \rho_{n} \kappa v_{y,n}
    \label{eq:notes015}
\end{equation}
In the $y$ direction we have,
\begin{equation}
  \label{eq:notes0161}
  F_{y} = \rho_{n} \kappa v_{x,n} -  \frac{\rho_{n}^{2}\kappa^{2}}{C} v_{y,n}
  = \rho_{n} \kappa v_{x,n} - \rho_{n} \kappa \frac{\beta}{\beta'} v_{y,n}.
\end{equation}
In the inertial frame where ${\bf v}_{c} \neq 0$ the force per unit length is:
\begin{align}
{\bf F}_{n} & = \rho_{n} {\bf \kappa} \times ({\bf v}_{n} - {\bf v}_{c})
- \frac{\rho_{n}^{2}\kappa^{2}}{C} ({\bf v}_n \cdot \hat{{\bf y}}) \hat{{\bf y}} \nonumber \\
            & = \rho_{n} {\bf \kappa} \times ({\bf v}_{n} - {\bf v}_{c})
            - \rho_{n} \kappa \frac{\beta}{\beta'}({\bf v}_n \cdot \hat{{\bf y}}) \hat{{\bf y}}
\end{align}
which clearly does not have the symmetries of Eq.(\ref{eq:MutualFriction}).

Assuming an array of vortices of constant density $n_{v}$ and expressing the velocities in terms of the angular velocity lag
$\omega$, we find the force density on the neutron superfluid to be
\begin{align}
  F_{x,n} &= - \rho_{n} (\kappa n_{v})\; \omega x \\
  F_{y,n} &= - \rho_{n} (\kappa n_{v})\; \omega y - \frac{\beta}{\beta'} \rho_{n} (\kappa n_{v})\; \omega x.
\end{align}

We should also note that the form of the force in (\ref{eq:notes015}) and (\ref{eq:notes0161}) in the limit of  strong damping by
electron scattering ($C \to \infty$) is the same as in the case with no magnetic flux lines.
In this strong coupling limit the vortices are locked into the charged component (proton superconductor plus normal electrons)
whether the magnetic flux lines are present or not and so the vortex dynamics are not modified by the flux lines.

\subsection{Macroscopic Dynamics}

Up to this point we have derived the macroscopic dynamics from considerations of vortex currents, as in Eq.(\ref{eq:spinDownSimpleInt}).
The dynamics of our two component system can also be formulated in terms of torques.
In this section, for illustrative purposes, we shall present both derivations.

The equations of motion in a frame rotating with the proton fluid will be given by
\begin{equation}
  \label{eq:EqnMotLX}
  I_{X} \dot{\bf \Omega}_{X} + I_{X} {\bf \Omega}_{c} \times {\bf \Omega}_{X} = \int_{Star}  {\bf r}\times {\bf F}_{X} d V
\end{equation}
where $X$ denotes either of $n$ and $c$.

For the neutron fluid component we find
\begin{equation}
  I_{n} \dot{\bf \Omega}_{n} + I_{n} {\bf \Omega}_{c} \times {\bf \Omega}_{n}
  = - 2 \rho_{n} \Omega_{n} \omega \int_{Star}\left\{ x {\bf r} \times \hat{{\bf x}}
  + {\bf r} \times \hat{{\bf y}} \left( y + \frac{\beta}{\beta'} x \right) \right\} dV
\end{equation}
Most of the integral vanishes by symmetry so we are left with
\begin{equation}
  I_{n} \dot{\bf \Omega}_{n} = - 2 \rho_{n} \Omega_{n} \omega \int_{Star} \frac{\beta}{\beta'} x^{2} \hat{{\bf z}}\; dV
\end{equation}
For a spherical star of constant density
\begin{equation}
  \rho_{n} \int_{Star} x^{2} dV = \frac{1}{5} \rho_{n} V R^{2} = \frac{I_{n}}{2}
\end{equation}
where $I_{n}$ is the total moment of inertia of the neutron fluid.
Taking the angular velocity vectors initially in the $\hat{{\bf z}}$ direction,
only the magnitudes of $\Omega_{X}$ will change, according to
\begin{align}
\label{eq:obsSpinDownN}
  I_{n} \dot{\Omega}_{n} &=  - I_{n} \Omega_{n} \omega \frac{\beta}{\beta'}  \\
\label{eq:obsSpinDownP}
  I_{c} \dot{\Omega}_{c} &= I_{n}\Omega_{n} \omega \frac{\beta}{\beta'}.
\end{align}

Alternatively, Equation (\ref{eq:obsSpinDownN}) can be derived using Eq.(\ref{eq:spinDownSimpleInt}),
the relation between spindown rate and vortex current.
In the present case
\begin{equation}
v_r = v_{x,L} \cos(\phi)= \frac{\rho_{n} \kappa}{C} v_{y,n} \cos(\phi) = \frac{\rho_{n} \kappa}{C} r \omega \cos^2(\phi),
\end{equation}
where we have used Eq.(\ref{eq:notes015}), and substituted $v_{y,n} = r \omega \cos(\phi)$.
The integration of radially outward vortex current leads to Eq. (\ref{eq:obsSpinDownN}), in the form
\begin{equation}
\dot{\Omega}_{n} = - \frac{\beta}{\beta'} \Omega_{n} \omega
= - \frac{\rho_{n}\kappa}{C}\Omega_{n}\omega \equiv - \frac{\omega}{\tau_{n,pin}}
\end{equation}
with
\begin{equation}
\tau_{n,pin} = \frac{C}{\rho_{n}\kappa\Omega_{n}} .
\end{equation}
In the presence of an external torque $N_{ext}$ aligned with the angular velocity vectors, Eq. (\ref{eq:obsSpinDownP}) becomes
\begin{equation}
\label{eq:totalSpinDown}
  I_{c} \dot{\Omega}_{c} =  - I_{n}\dot{\Omega}_{n} + N_{ext},
\end{equation}
so that the equation for the lag is
\begin{equation}
    \label{eq:something01}
  \dot{\omega} = - \Omega_{n} \omega \frac{\beta}{\beta'} \frac{I}{I_{c}} - \frac{N_{ext}}{I_{c}}.
\end{equation}
Observed rotation rates of neutron stars have variations of at most a fraction $\Delta\Omega_{c} / \Omega_{c} \sim 10^{-6}$
even in the largest glitches. The lag $\omega$ must therefore be small compared to $\Omega_{c}$ and $\Omega_{n}$, and
(\ref{eq:something01}) can be solved with $\Omega_{n}$ set to a constant representative value $\Omega_{0}$. The solution is
\begin{equation}
\label{eq:omegaObsSol}
  \omega (t) = (\omega(0)- \omega_{\infty}) \exp\left(\;- \frac{t}{\tau_{pin}} \right) + \omega_{\infty}
\end{equation}
which is just Eq.(\ref{eq:lagSolSimp}) with the relaxation time $\tau_{pin}$ appropriate for the present situation.
The dynamical timescale $\tau_{pin}$ over which the two fluids couple is related to the timescale $\tau_{n,pin}$,
with the generic relation
\begin{equation}
\tau_{pin} = \frac{\rho_{c}}{\rho}\tau_{n,pin}
\end{equation}
leading to
\begin{equation}
\tau_{pin} = \frac{\rho_{c}}{\rho \Omega_{0}} \frac{\beta'}{\beta} =
\frac{\rho_{c}}{\rho \Omega_{0}} \frac{C}{\rho_{n} \kappa}
\end{equation}
and $\omega_{\infty} = |\dot{\Omega}_{\infty}|\tau_{n,pin}$.
The solutions for $\Omega_{n}(t)$ and $\Omega_{c}(t)$ are obtained from Eqs. (\ref{eq:totalSpinDown}) and (\ref{eq:omegaObsSol}).
Thus the introduction of flux line - vortex line pinning as an absolute barrier to vortex motion in one direction
reduces the coupling time by a factor
\begin{equation}
    \frac{\tau_{pin}}{\tau} = \frac{\tau_{n,pin}}{\tau_{n}}
    = 2 \beta' \cong 2 \left( \frac{C}{\rho_{n}\kappa} \right)^2 \sim O(3.2 \times 10^{-7} - 8 \times 10^{-10}).
\label{eq:reducedCouplingTime}
\end{equation}

\section{vortex creep}

\subsection{Vortex Creep in the Axially Symmetric Case}

At finite temperature vortex lines can pass (creep) through flux tubes via thermal fluctuations.
We shall model the resulting vortex creep velocity in the $y$ direction by generalizing the model developed for the neutron star crust
superfluid where vortices creep against an isotropic distribution of point pinning centres (Alpar et al 1984a).

We first review this familiar model to set the stage for anisotropic creep against flux lines.
For a vortex line to unpin there is an energy barrier of size $E_{p}$ to overcome.
The rate of spindown of a superfluid is determined by the vortex velocity in the radially outward direction from the rotation axis,
as we saw in Eq. (\ref{eq:spinDownSimple}). The creep velocity in the radial direction is
\begin{equation}
    v_{r} = v_{0} \left[ \exp\left(\;-\frac{E_{p-out}}{kT}\right) - \exp\left(\;-\frac{E_{p-in}}{kT}\right) \right]
\end{equation}
where $v_{0}$ is the average trial velocity against pinning potentials, $E_{p-out}$ and $E_{p-in}$ are
the pinning energy barriers for radially outward and inward motion, respectively.
If there is no lag between the two fluid constituents, $\omega = 0$, then there is no difference in pinning energy barriers
for vortices creeping outwards or inwards; $E_{p-out} = E_{p-in}$ and so $v_{r} = 0$. When the star is spinning down,
driven by an external torque,
the crust and charged constituent is rotating faster than the neutron superfluid ($\omega > 0$).
Vortex motion in the outwards direction will reduce this lag and so is energetically favourable.

A pinned vortex line is dragged along with the crust and charged fluid. A velocity difference will be set up between
the line and the neutron superfluid. According to the Magnus equation, Eq.(\ref{eq:MagnusForce}), a radially inward force must be applied
on the vortex line by the pinning centres. The pinning energy barrier in the outward direction is reduced by an amount
$\Delta E$ proportional to the lag $\omega$, while the barrier to motions in the radially inward direction is enhanced by the same amount.
The pinning barrier $E_{p}$ can sustain a maximum lag $\omega_{crit}$; there is not enough pinning force to keep vortices
pinned at higher lags:
\begin{equation}
\label{eq:pinEnergCrit}
    E_{p} = b \xi F_{critical} = b \xi \rho_{n} \kappa \omega_{crit} r
\end{equation}
where $b$ is the distance between pinning sites and $\xi$, the coherence length, is the radius of the vortex core,
the distance scale over which the pinning force acts. The second equation relates the critical pinning force per
unit length of vortex line to the critical lag through the Magnus equation. At the actual lag $\omega < \omega_{crit}$
the correction $\Delta E$ to the pinning energy $E_{p}$ is
\begin{equation}
    \Delta E = b \xi F = b \xi \rho_{n} \kappa \omega r.
\end{equation}
The energy barrier in the outwards direction is given by
\begin{align}
    E_{p-out} &= E_{p} - \Delta E \\
              &= E_{p} \left( 1 - \frac{\omega}{\omega_{crit}}\right).
\end{align}
Vortex motion in the radially inwards direction faces an energy barrier enhanced by $\Delta E$, so that
\begin{equation}
    E_{p-in} = E_{p} \left( 1 + \frac{\omega}{\omega_{crit}}\right).
\end{equation}
This finally gives the radial creep velocity
\begin{equation}
\label{eqn:creepVelBasic}
  v_{r} = 2 v_{0} \exp \left( -\frac{E_{p}}{kT} \right) \sinh \left( \frac{E_{p}}{kT} \frac{\omega}{\omega_{crit}} \right) .
\end{equation}
Substituting the radial creep velocity in the equation for superfluid spindown, Eq.(\ref{eq:spinDownSimple}) completes the creep model
for the case of an isotropic distribution of point pinning centres.
There will be no creep in the azimuthal direction.

\subsection{Vortex Creep Across Flux Lines}

In the presence of flux lines, vortex motion in the $y$ direction will be determined by creep through the flux lines.
We model vortex motion in the $y$ direction in terms of creep alone, $v_{L,y} = v_{creep}(\omega)$,
ignoring the effects of drag forces. This approach is appropriate in the case of weak coupling drag forces, and is justified
for the drag forces given in Eqs.(\ref{eq:tauvortP}) and (\ref{eq:weakcoupling}). Strictly speaking,
a vortex line moving in the $y$ direction experiences drag forces between encounters with flux lines. For weak enough drag,
the governing rate (bottleneck) for motion in the $y$ direction is given by the very slow rate of creep across the flux lines,
so that the effect of weak coupling drag forces can be neglected. Strong coupling, with drag and creep having comparable effects
on the motion of vortex lines in the $y$ direction requires a more complicated treatment which will not be adressed in this paper.

To find the appropriate expression for $v_{creep}$ we must recalculate $E_{p-out}$ and $E_{p-in}$ or
rather in this case the effective pinning energies of the positive and negative directions along $y$ ($E_{p+}$ and $E_{p-}$ respectively).
Again any bias must be due to vortex motion in the radially outward  or inward direction. This is taken into account
by introducing a factor $\hat{\bf r} \cdot \hat{\bf y}=\sin (\phi)$:
\begin{align}
    E_{p+} &= E_{p} \left[ 1 - \frac{\omega}{\omega_{crit}} \sin (\phi) \right] \nonumber \\
    E_{p-} &= E_{p} \left[ 1 + \frac{\omega}{\omega_{crit}} \sin (\phi) \right].
\end{align}
The vortex lines' velocity in the $y$ direction, due to vortex creep is thus
\begin{equation}
\label{eqn:creepVelFluxlines}
v_{L,y} =  v_{creep}(\omega \sin(\phi)) =
2 v_{0} \exp \left( -\frac{E_{p}}{kT} \right) \sinh \left( \frac{E_{p}}{kT} \frac{\omega \sin(\phi)}{\omega_{crit}} \right) .
\end{equation}

In the $x$ direction, vortices move parallel to the flux lines, at velocity $v_{L,x}$
encountering only the drag force due to scattering of charged particles off the magnetized vortex core.
The components of the Magnus equation Eq.(1) with anisotropy in the rotating frame of the charged component $c$,
analogous to Eqs.(\ref{eq:notes013}) and (\ref{eq:carenotes014}), are
\begin{align}
  C v_{L,x} &= \rho_{n} \kappa \left( v_{n,y} - v_{creep} \right) = \rho_{n} \kappa \left( \omega r \cos (\phi) - v_{creep} \right)
  \label{eq:pincreep1}\\
  F_{y}(\omega) &=  \rho_{n} \kappa \left( v_{n,x} - v_{L,x} \right) = \rho_{n} \kappa \left( - \omega r \sin (\phi) - v_{L,x} \right).
  \label{eq:pincreep2}
\end{align}
Since the vortex line velocity in the $y$ direction can, in principle,
be a nonlinear function of $\omega$, the physical force per unit length $F_{y, creep}$ representing the
interaction of the creeping vortex with the flux lines can also be a nonlinear function of $\omega$.

With these two different vortex line velocities in the $x$ and $y$ directions, the equation of superfluid spindown,
Eq. (\ref{eq:spinDownSimpleInt}) becomes
\begin{equation}
\dot{\Omega}_{n} = - \frac{\Omega_{n}}{\pi r^{2}} \oint \left[ v_{L,x} \cos (\phi)  + v_{L,y} \sin (\phi) \right] r d \phi.
\end{equation}
Substitution from Eqns. (\ref{eqn:creepVelFluxlines}) and (\ref{eq:pincreep1}) leads to
\begin{equation}
\dot{\Omega}_{n} = - \frac{\Omega_{n}}{\pi r^{2}}
\oint \left[\frac{\rho_{n} \kappa}{C} \left( \omega r cos^2 (\phi) - v_{creep}(\omega \sin (\phi)) \cos (\phi) \right)
+ v_{creep}(\omega \sin (\phi)) \sin (\phi) \right] r d \phi.
\end{equation}
The second term clearly integrates to zero. Evaluating the integral in the first term, our equation of motion is now given by
\begin{equation}
  \dot{\Omega}_{n} = - \frac{\rho_{n} \kappa}{C} \Omega_{n} \omega - \frac{2 \Omega_{n} v_{0}}{\pi r}\exp \left( -\frac{E_{p}}{kT} \right)
  \oint \sinh \left[ \frac{E_{p}}{k T} \frac{\omega}{\omega_{crit}} \sin (\phi) \right] \sin (\phi) d \phi
    \label{eq:VelVortexPolar}
\end{equation}

Both of these terms have the same sign, and so taking the first term alone will give us an upper limit on the lag $\omega$.
Taking the Vela pulsar as an example, we have $\dot{\Omega} \approx - 10^{-10}$ rad s$^{-2}$,
$\Omega \approx 70$ rad s$^{-1}$. The range of values of $C/(\rho_n \kappa)$ quoted in Eq.(\ref{eq:weakcoupling})
imply that the lag will be less than $\sim 6 \times 10^{-16}$ rad s$^{-1}$.
This bound implies that the creep term is also linear in the lag $\omega$ since the argument of sinh is very small,
as can be seen using Eq. (\ref{eq:pinEnergCrit}), with a temperature $T = 10^8 K$ and typical parameter values for the
neutron star core.
Eq.(\ref{eq:VelVortexPolar}) becomes
\begin{equation}
  \label{eq:creepTime}
\dot{\Omega}_{n} = - \frac{\rho_{n} \kappa}{C} \Omega_{0} \omega
- \frac{2 \Omega_{0} v_{0}}{ R} \exp \left( - \frac{E_{p}}{kT} \right) \frac{E_{p}}{k T} \frac{\omega}{\omega_{crit}}
\end{equation}
where we have substituted a constant fiducial value $\Omega_{0}$ for the slowly varying $\Omega_{n}$, and taken $R = 10^6$ cm
to represent the radius $r$.
Introducing a relaxation time
\begin{equation}
\label{eq:creepTime1}
\tau_{n,creep} \equiv \frac{R}{2 \Omega_{n} v_{0}} \exp \left( \frac{E_{p}}{kT} \right) \frac{k T \omega_{crit}}{E_{p}}
\end{equation}
and an effective relaxation time
\begin{equation}
\frac{1}{\tau_{n,eff}} \equiv \frac{1}{\tau_{n,pin}} + \frac{1}{\tau_{n,creep}}
\end{equation}
Eq. (\ref{eq:creepTime}) can be written as
\begin{equation}
  \label{eq:effTime}
\dot{\Omega}_{n} = - \frac{\omega}{\tau_{n,eff}}.
\end{equation}
As in the earlier sections, treating this coupling together with the angular momentum evolution, Eq.(\ref{eq:totalSpinDown}), leads to
solutions of the form given in Eq. (\ref{eq:omegaObsSol}) with the dynamical relaxation time
\begin{equation}
  \label{eq:effTime2}
{\tau_{eff}} = \frac{\rho_{c}}{\rho}{\tau_{n,eff}} = \frac{\tau_{pin}\tau_{creep}}{\tau_{pin} + \tau_{creep}}
\end{equation}
where the dynamical relaxation time for creep (in the $y$ direction) is
\begin{equation}
  \label{eq:creepTime2}
{\tau_{creep}} = \frac{\rho_{c}}{\rho}{\tau_{n,creep}}.
\end{equation}
The effective relaxation time is the shorter of $\tau_{pin}$ and $\tau_{creep}$. The ratio of these relaxation times is
\begin{equation}
\frac{\tau_{creep}}{\tau_{pin}}
= \frac{ R}{2 \Omega_{n} v_{0}} \exp \left( \frac{E_{p}}{kT} \right) \frac{k T \omega_{crit}}{E_{p}} \frac{\rho_{n}\kappa\Omega_{n}}{C}.
\end{equation}
We take $T = 10^8 K, kT \approx 10^{-2}$ MeV, a typical density
$\rho = 3 \times 10^{14}$ g cm$^{-3}$, $v_{0} = 10^7$ cm s$^{-1}$, $R = 10^6$ cm,
the coherence length $\xi = 10^{-11}$ cm and the distance between pinning junctions (flux line spacing)
$b \cong 5 \times 10^{-10} (B_{12})^{-1/2}$, where $B_{12}$ denotes the magnetic field in the neutron star core in units of $10^{12}$ G.
We find that $\tau_{creep}$ is the longer timescale unless $E_{p}/kT < 9$. With estimates of the pinning
energy $E_p \sim 0.1 - 1$ MeV (Ruderman, Zhu \& Chen 1998, Srinivasan et al 1990) and $kT \approx 10^{-2}$ MeV, the creep rate is likely to be negligible,
and the core superfluid remains in the very tight coupling with the charged component imposed by the presence of the flux lines.
This result depends only logarithmically on the parameter values we have used.

If the pinning energy is closer to the thermal energy ($E_{p}/ k T < 5 $) then
the creep process determines the relaxation of the core, $\tau_{eff} \cong \tau_{creep}$, somewhat alleviating the almost rigid coupling
to the charged component inferred in Eq.(\ref{eq:reducedCouplingTime}).

\section{toroidal field}

Recent stability simulations of the type of magnetic fields expected in neutron stars (Braithwaite 2008)
suggest the presence of a large toroidal contribution to the internal magnetic field.
In this case the flux lines prevent radially outwards motion of the superfluid vortices in all directions, so that any
spin down of the superfluid must be due to creep. Because of azimuthal symmetry, the macroscopic dynamics will in fact be the
same as in the creep model for
the crust superfluid. If toroidal flux lines coexist with the uniform array of flux lines aligned in the $x$ direction,
reconnection will lead to a toroidal field;
in any case there is no ``easy'' direction with vortices moving outwards against drag forces only.
We therefore assume that the toroidal field is confined to an
equatorial belt surrounding but not overlapping with the core region with poloidal field lines
(here poloidal means a uniform parallel array of flux lines).  The dominant physical process
constraining the radial vortex current is creep against the torroidal flux lines.
The radially outward vortex current will have its dynamics
characterized by the {\em local} values of the lag $\omega$. In the region of toroidal flux lines the lag $\omega$
may have quite different values from the rest of the core.

The equation of motion is
\begin{equation}
\dot{\Omega}_{n} = - \frac{4 \Omega_{0}v_{0}}{R} \exp \left( -\frac{E_{p}}{k T} \right)
\sinh \left( \frac{E_{p}}{k T}\frac{\omega}{\omega_{crit}}\right).
\label{eq:creeptoroidal}
\end{equation}
Vortex creep can operate in the linear or nonlinear regimes (Alpar, Cheng \& Pines 89), depending on a comparison of the
spin down rate determined by the external torque and the efficiency of thermal creep, as reflected in the magnitude
of the sinh function
\begin{equation}
\sinh \left( \frac{E_{p}}{k T}\frac{\omega}{\omega_{crit}} \right) = \frac{4 \Omega_{0}v_{0}}{|\dot{\Omega}_{n}| R}
\exp \left( \frac{E_{p}}{k T} \right)
= \frac{8 \tau_{spindown} v_{0}}{R} \exp \left( \frac{E_{p}}{k T} \right)
\label{eq:sinh}
\end{equation}
where $\tau_{spindown} \equiv \Omega_{0}/2|\dot{\Omega}|$ is the characteristic spindown age of the pulsar.
For the Vela pulsar parameters, the estimate of the sinh is less than 1 if $E_{p}/k T < 31$, or $E_{p} < 0.3$ MeV,
if $T = 10^{8} \; {}^o$K.
In this case Eq.(\ref{eq:creeptoroidal}) is
linearized to give
\begin{equation}
\dot{\Omega}_{n} = - \frac{4 \Omega_{0}v_{0}}{R} \exp \left( -\frac{E_{p}}{k T} \right)
\frac{E_{p}}{k T}\frac{\omega}{\omega_{crit}} \equiv - \frac{2 \omega}{\tau_{n,creep}}
\equiv - \frac{ \omega}{\tau_{n,tor}}.
\label{eq:lincreeptoroidal}
\end{equation}
Thus, in the linear regime creep across toroidal flux lines the relaxation times $\tau_{n,tor}$ and $\tau_{tor}$
are half the corresponding relaxation times $\tau_{n,creep}$ and $\tau_{creep}$ for creep
across flux lines aligned in the $x$ direction.
This is easy to understand since creep operates in the $r$ direction, that is, both $x$ and $y$ motions are allowed,
while in the
latter case creep takes place only in the $y$ direction. The relaxation time of the toroidal region
has a sensitive dependence on $E_p/(kT)$ :
\begin{equation}
\tau_{tor} = \frac{\rho_c}{\rho}\tau_{n,tor}
= \frac{\rho_c}{\rho} \frac{ R}{4 \Omega_{n} v_{0}} \exp \left( \frac{E_{p}}{kT} \right)
\frac{k T \omega_{crit}}{E_{p}} \sim 10^{-8} T_8 \exp \left( \frac{E_{p}}{kT} \right) \textrm{s}
\end{equation}
Thus $\tau_{tor}\sim 3 \times 10^5 s $ for $E_{p}/k T = 31$, the limiting value for linear creep.
If $E_{p}/k T$ is somewhat smaller, the dynamical relaxation time is reduced exponentially;
for example $\tau_{tor} \sim < 2\times 10^{-4} $ s for $E_{p}/k T = 10$.

\section{conclusions}

If flux lines of the proton superconductor are not taken into account, the dynamical relaxation time for the core superfluid is
\begin{equation}
  \tau = \frac{\rho_c}{\rho}\tau_n = \frac{\rho_n}{\rho}\tau_v \cong \tau_v \cong (10-200) P.
\end{equation}
as given in Eqs. (\ref{eq:tauvortP}) and (\ref{eq:scaleTauNTau}), employing the results of (Alpar, Langer \& Sauls 1984).
The range of uncertainty corresponds to the uncertainty in the representative values of the effective proton mass in
the neutron star core.

We have found that the inclusion of the proton flux lines strengthens the coupling between
the neutron superfluid in the core and the charged component of the star, leading to substantially shorter dynamical coupling times.
When flux lines are included as absolute barriers in a poloidal like configuration,
the neutron superfluid spins down by the outward motion of the vortex lines in the ``easy" direction,
parallel to the flux lines, checked only by drag forces.
The dynamical coupling time is reduced by the factor $2 \beta' \cong 2 [C/(\rho_n \kappa)]^2$,
where $C$ is the coefficient of the drag force and $\rho_n \kappa$ is the Magnus inertial coefficient, yielding
\begin{equation}
    \frac{\tau_{pin}}{\tau} \sim 3.2 \times 10^{-7} - 8 \times 10^{-10}.
\end{equation}
Absolute pinning to poloidal flux lines forces the neutron superfluid into effective corotation with
the charged component and outer crust of the neutron star. Allowing vortex creep across the flux lines does not
effect this conclusion. For relevant neutron star core temperatures and
pinning energy estimates for flux line - vortex line junctions, the timescale for creep across the flux lines is typically
much longer than the timescale $\tau_{pin}$ of dynamical coupling achieved by vortex motion parallel
to the flux lines alone, so that the effective dynamical relaxation time with creep present is
not changed, $\tau_{eff} \cong \tau_{pin}$. Thus the presence of flux lines reduces dynamical coupling times
drastically and clamps the neutron star
core superfluid to the crust and charged component.

The likely toroidal arrangement of flux lines circumscribing the core superfluid does not allow any anisotropy
in vortex flow. All vortices have to creep
across the toroidal flux lines if they are to move out radially. If $E_{p}/k T$ is small enough
($E_{p}/k T < 31$ for Vela pulsar parameters, with a similar critical value for other pulsars) creep is
linear, the dynamical relaxation time is simply the creep relaxation time
$\tau_{tor} \sim 10^{-8} T_8 \exp \left( \frac{E_{p}}{kT} \right)$ s.
The toroidal region may well contain values of the pinning energy with $E_{p}/k T$ large enough for
the nonlinear creep regime. In the nonlinear regime, corresponding, for the Vela pulsar parameters, to $E_{p}/k T > 31$, or $E_{p} > 0.3$ MeV, if $T = 10^{8}$K,
The sinh function is approximated with $\sinh(x) = 1/2 \exp(x)$.
The resulting highly nonlinear dynamical model has been successfully
applied to the dynamics of the inner crust superfluid, and invoked to explain basic features of postglitch and interglitch
response in pulsars (Alpar et al 1984b). With its location at the boundary between the neutron star core and superfluid inner crust,
the region with toroidal flux lines could have a moment of inertia fraction $I_p/I \sim O(10^{-2})$, comparable to the
fractional moment of inertia of the superfluid inner crust. The steady state lag $\omega_{\infty} \cong \omega_{crit}$ and
the relaxation times for nonlinear creep in the toroidal region may also be similar to the lag and relaxation times of the inner
crust superfluid. The toroidal region will then have postglitch response signatures similar to those of the inner crust superfluid, and
will be responsible for some of the nonlinear response previously attributed to the inner crust superfluid. This makes
it impossible to draw unambiguous lower limits on the moment of inertia fraction of the inner crust superfluid to use
for constraints on the neutron star structure and equation of state.

Nonlinear creep implies a much larger value of the steady state lag $\omega_{\infty}$ in the toroidal region
than the typical lags in the linear regime, which we found to be extremely small in the neutron star core with
its poloidal (uniform) array of flux lines.
This change in $\omega$ means a sharp increase in $\Omega_n$, and therefore a large density of neutron vortex lines in the boundary
between the core superfluid and the toroidal region. This may well be an axially symmetric vortex reservoir to provide the source
of glitches when vortex lines are unpinned catastrophically. Thus the toroidal region could become an alternative,
or a complement to the inner crust superfluid as a source of glitches as well as being a prominent agent of postglitch relaxation.


\begin{thebibliography}{12}

\bibitem{AlpAndPinSha84a}
Alpar, M.A., Anderson, P.W., Pines, D., Shaham, J., ApJ. {\bf 276} 325 (1984a)

\bibitem{AlpAndPinSha84b}
Alpar, M.A., Anderson, P.W., Pines, D., Shaham, J., ApJ. {\bf 278} 791 (1984b)

\bibitem{Alpetal93}
Alpar, M.A., Chau, H.F., Cheng, K.S. \& Pines, D., ApJ. {\bf 409} 345 (1993)

\bibitem{AlparChengPines89}
Alpar, M.A., Cheng, K.S. \& Pines, D., ApJ. {\bf 346} 823 (1989)

\bibitem{AlpLanSau84}
Alpar, M.A., Langer, S.A. \& Sauls, J.A., ApJ. {\bf 282} 533 (1984)

\bibitem{AlpSau88}
Alpar, M.A., Sauls, J.A., ApJ. {\bf 327} 723 (1988)

\bibitem{AndSidCom06}
Andersson, N., Sidery, T. \& Comer, G.L., Mon.Not.Roy.Astron.Soc. {\bf 368} 162 (2006)

\bibitem{BaldoDucoin09}
Baldo, M. \& Ducoin, C., Physical Review C {\bf 79} 035801 (2009)

\bibitem{Braith08}
Braithwaite, J., arXiv:0810.1049 (2008)

\bibitem{ChamHaen06}
Chamel, N. \& Haensel, P., Phys.Rev.C {\bf 73} 045802 (2006)

\bibitem{Dod2007}
Dodson, R., Lewis, D. \& McCulloch, P., ApSS {\bf 308} 585 (2007)

\bibitem{RudermanZhuChen}
Ruderman, M.A., Zhu, T. \& Chen, K., Ap.J. {\bf 492} 267 (1998)

\bibitem{Sauls88}
Sauls, J.A., in "Timing Neutron Stars", H. \"{O}gelman \& E.P.J. van den Heuvel, eds., Kluwer (1989)

\bibitem{SedSed}
Sedrakian, A.D., Sedrakian, D.M., Ap.J. {\bf 447} 305 (1995)

\bibitem{Sidery09}
Sidery, T., in preparation (2009)

\bibitem{Srinietal90}
Srinivasan, G., Bhattacharya, D., Muslimov, A.G., \& Tsygan, A.J., Current Science {\bf 59} 31 (1990)

\end{thebibliography}
\end{document}